\documentclass{article}

\usepackage[final]{nips_2016}
\usepackage{graphicx}

\usepackage[utf8]{inputenc} 
\usepackage[T1]{fontenc}    
\usepackage{hyperref}       
\usepackage{url}            
\usepackage{booktabs}       
\usepackage{amsfonts}       
\usepackage{nicefrac}       
\usepackage{microtype}      

\title{Multi-task Learning in the Computerized Diagnosis of Breast Cancer on DCE-MRIs}

\author{
  Natalia Antropova, Benjamin Huynh, Maryellen Giger \\
  Department of  Radiology\\
  The University of Chicago\\
  Chicago, IL 60637 \\
  \texttt{antropova@uchicago.edu} \\
}

\begin{document}
\maketitle

\begin{abstract}
Hand-crafted features extracted from dynamic contrast-enhanced magnetic resonance images (DCE-MRIs) have shown strong predictive abilities in characterization of breast lesions. However, heterogeneity across medical image datasets hinders the generalizability of these features. One of the sources of the heterogeneity is the variation of MR scanner magnet strength, which has a strong influence on image quality, leading to variations in the extracted image features. Thus, statistical decision algorithms need to account for such data heterogeneity. Despite the variations, we hypothesize that there exist underlying relationships between the features extracted from the datasets acquired with different magnet strength MR scanners. We compared the use of a multi-task learning (MTL) method that incorporates those relationships during the classifier training to support vector machines run on a merged dataset that includes cases with various MRI strength images. As a result, higher predictive power is achieved with the MTL method.
\end{abstract}

\section{Introduction}

Computer aided diagnosis (CADx) has been widely used for characterization of breast lesions based on dynamic contrast-enhanced magnetic resonance images (DCE-MRIs). Traditional CADx systems automatically segment a lesion from the neighboring background and subsequently extract intuitive hand-crafted features that describe its size, shape, enhancement, or kinetics \textsuperscript{1-5}. Once the features are extracted, they are used to build classification models to make diagnostic decisions (e.g. likelihood of malignancy, molecular characteristics). Despite the success of CADx predictive models, DCE-MRI CADx faces many challenges, one being the heterogeneity within the MR image data. One of the sources of the data instability comes from the variability of MR magnet strength. Currently, clinical breast imaging is performed with 1.5 Tesla (T) or 3 T MR scanners. The magnet strength has an effect on image quality, with lower strength scanners producing images with higher noise and lower resolution, leading to less resolved anatomical details \textsuperscript{6}. Thus, CADx features may vary with the MRI magnet strength.  

Unfortunately, currently available medical image datasets are of limited sizes. Separating the DCE-MRI data based on magnet strength is undesirable since it would further decrease the dataset sizes. The goal of this work is to search for a generalizable predictive model that will allow for incorporation of both 1.5T and 3T data. We hypothesize that despite the described differences, there exist underlying relationships between features extracted from images acquired with 1.5T and 3T scanners. Based on this assumption, we explore multi-task learning (MTL) to generalize discrimination of malignant and benign lesions over different feature domains. The approach assumes that there exists relatedness between the different-domain features, which is utilized to improve the classification performance. 

\section{Materials and methods}

\subsection{Database}
Assessment of the MTL method was performed on 1.5T and 3T DCE-MRI datasets of 447 and 193 breast lesions, respectively (Table~\ref{dataset-table}). The DCE-MR images were acquired at the University of Chicago Medical Center over a span of 6 years from 2006 to 2012 on either 1.5T or 3T Philips scanners with T1-weighted spoiled gradient sequence. Each lesion was characterized as benign or malignant based on pathology from biopsy. Examples of 1.5T and 3T images of cancerous lesions are shown in Figure 1. 

\begin{table}[ht]
	\caption{The number of cases in 1.5T, 3T and merged (1.5T+3T) datasets. The table includes image resolution in each dataset. The resolution varies drastically between specific-magnet-strength datasets.} 
	\centering 
	\label{dataset-table}
	\begin{tabular}{c c c c} 
		\toprule 
		 & 1.5T Dataset & 3T Dataset & Merged Dataset \\ [0.5ex] 
		\midrule 
		Benign & 149  & 42 &	191 \\ 
		Malignant & 298 & 151 & 449 \\
		Total & 447  & 193 & 640 \\
		\midrule
		Resolution  (mm) & 0.72 – 1.35 &	0.54 – 0.59	& 0.54 – 1.35  \\
		\bottomrule
	\end{tabular}
	\label{table:nonlin} 
\end{table}

\begin{figure}[h]
	\begin{minipage}[b]{1.00\linewidth}
		\centering
			\includegraphics[width=0.48\textwidth, height =2.8 in,clip=TRUE]{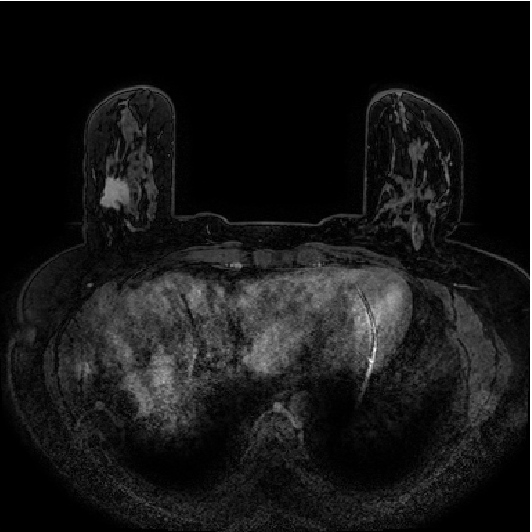}
			\includegraphics[width=0.48\textwidth, height =2.8 in,clip=TRUE]{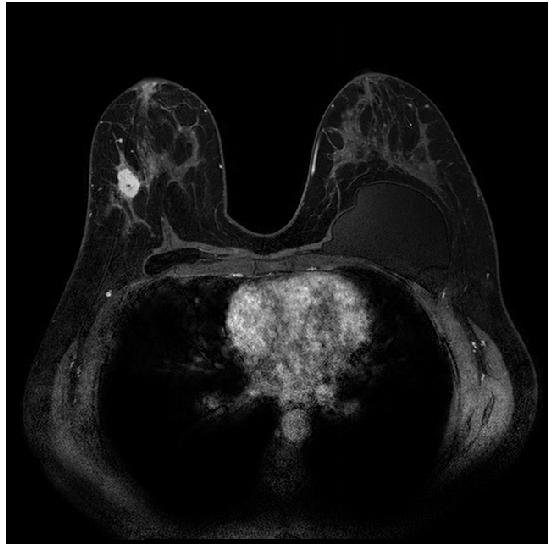}
		\caption{DCE-MR images of cancerous lesion acquired on 1.5 T scanner with  0.964 mm resolution (left) and on 3 T scanner with  0.547 mm resolution (right).}
	\end{minipage}
\end{figure}

\subsection{Hand-crafted CADx features}
Lesion classification was performed with 38 hand-crafted features, extracted from the lesion segmentations on DCE-MRIs \textsuperscript{1-5}.The hand-crafted features characterize lesions in terms of its size, shape, morphology, enhancement texture, kinetics, or kinetics variance. Features that describe shape, marginal characteristics, and spatial enhancement patterns can especially be influenced by the image resolution and noise.

\subsection{Classification algorithms:  MTL and SVM}
We utilized a multi-task relationship learning approach described by Zhang and Yeung, which models relationships between the tasks  to discriminate benign and malignant lesions\textsuperscript{7}. The approach uses a prior on {\bf W} (the matrix of the model weights), which is modeled with a matrix-variate normal distribution with the probability density function $p ({\bf W} \vert M, I, \Omega)$. Here, covariance matrix $\Omega$ incorporates the relationships between the task-specific model weights. We compared classification performances of the multi-task algorithm to our previously used support vector machines (SVM) in the task of distinguishing between benign and malignant lesions. The SVM was run on the merged MRI dataset, obtained by combining 1.5T and 3T datasets. The hyperparameters of both algorithms were optimized via grid search with nested five-fold cross-validation. The evaluation of the algorithms was performed with the ROC analysis \textsuperscript{8}.

\begin{table}[t]
	\caption{Performance metrics values for classification of the merged dataset by MTL and SVM classifiers. For a given sensitivity value, the MTL method outperforms the SVM method. }
	\label{sample-table}
	\centering
	\begin{tabular}{c c c p{1cm} c c p{1cm} p{1cm} c c p{1cm}}
		\toprule
		Sensitivity & \multicolumn{2}{c}{Specificity} & \multicolumn{4}{c}{Positive Predictive Value (PPV)} & \multicolumn{4}{c}{Negative Predictive Value (NPV)}\\
		\midrule
		& {\it MTL} &  {\it SVM} & & {\it MTL} & {\it SVM} & & &  {\it MTL} & {\it SVM} &  \\
		0.900 &	0.689 &	0.652 & &	0.552 &	0.524 & & & 	0.942 &	0.939 &  \\
		0.910 &	0.667 &	0.629 & &	0.537 &	0.511 & & & 	0.946 &	0.943 &  \\
		0.921 &	0.642 &	0.607 & & 0.522 &	0.499 & & & 	0.950 &	0.947 &  \\
		0.930 &	0.617 &	0.583 & &	0.508 &	0.487 & & & 	0.954 &	0.953  & \\
		0.940 &	0.587 &	0.555 & &	0.492 &	0.473 & & & 	0.958 &	0.958  & \\
		0.950 &	0.551 &	0.522 & &	0.474 &	0.457 & & & 	0.963 &	0.959  & \\
		0.960 &	0.507 &	0.481 & &	0.453 &	0.440 & & & 	0.967 &	0.966 &  \\
		0.970 &	0.455 &	0.434 & &	0.431 &	0.421 & & & 	0.973 &	0.971 &  \\
		0.980 &	0.386 &	0.370 & &	0.404 &	0.398 & & & 	0.978 &	0.977  & \\
		0.990 &	0.286 &	0.279 & &	0.371 &	0.369 & & & 	0.985 &	0.984  & \\
		
		\bottomrule
	\end{tabular}
	
\end{table}

\section{Results and discussion}

Table~\ref{sample-table} summarizes performance metrics for the MTL and SVM classifiers, while varying the decision boundary. Compared to the SVM classification, the MTL model results in higher positive (PPV) and negative (NPV) predictive values and therefore, it has more correct than incorrect identifications of benign and malignant lesions. Specificity values are higher for the MTL model, indicating reduction of false positive predictions.

We conclude that MTL has higher performance than SVM classifier with the resulting cross-validated AUC values of $AUC_{MTL}$= 0.90 (standard error (se) = 0.01) and $AUC_{SVM}$ = 0.88 (se = 0.01), for MTL and SVM classifiers, respectively. Figure 2 shows the ROC curves for the classification.

\begin{figure}[h]
	\begin{minipage}[b]{1.00\linewidth}
		\centering
		\centerline{
			\includegraphics[width=2.5in, height =2.5 in,clip=TRUE]{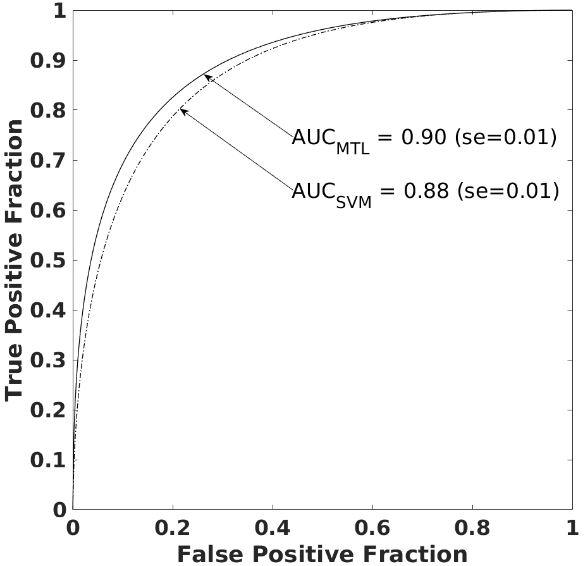}
		}
		\caption{ROC curves for the MTL and SVM classifiers in the task of distinguishing malignant and benign lesions. The classification is performed on the merged dataset (1.5T +3T).}
	\end{minipage}
\end{figure}

\clearpage

\section*{References}

[1] Gilhuijs, K.G., Giger ML,\ \& Bick U.\ (1998) Computerized analysis of breast lesions in three dimensions using dynamic magnetic-resonance imaging. {\it Medical Physics} {\bf 25}(9):1647–1654. 

[2] 	Gibbs, P., \ \& Turnbull, L.W.\ (2003) Textural analysis of contrast-enhanced MR images of the breast. {\it Magnetic Resonance in Medicine}  {\bf 50}(1):92–8. 

[3]	Chen, W., Giger, M.L., Li H., Bick, U.,\ \& Newstead, G.M.\ (2007) Volumetric texture analysis of breast lesions on contrast-enhanced magnetic resonance images. {\it Magnetic Resonance in Medicine} {\bf 58}(3):562–71. 

[4] Bhooshan, N., Giger, M.L., Jansen, S.A., Li H., Lan L.,\ \& Newstead, G.M.\ (2010) Cancerous breast lesions on dynamic contrast-enhanced MR images: computerized characterization for image-based prognostic markers. {\it Radiology} {\bf 254}(3):680–90. 
	
[5] Chen W., Giger, M.L., Bick, U., \ \& Newstead, G.M.\ (2006) Automatic identification and classification of characteristic kinetic curves of breast lesions on DCE-MRI. {\it Medical Physics} {\bf 33}(8):2878–87. 

[6] Rahbar, H., Partridge, S. C., DeMartini, W. B., Thursten, B.,\ \& Lehman, C. D. \ (2013). Clinical and technical considerations for high quality breast MRI at 3 Tesla. {\it Journal of Magnetic Resonance Imaging} {\bf 37}(4): 778-790.

[7] Zhang, Y.,  \ \& Yeung, D. Y. \ (2012) A convex formulation for learning task relationships in multi-task learning. {\it arXiv preprint arXiv}:1203.3536.

[8] Metz, C.,\ \& Pan, X. \ (1992) “Proper” Binormal ROC Curves: Theory and Maximum-Likelihood Estimation. {\it Journal of Mathematical Psychology} {\bf 43}(1):1–33. 
\end{document}